\documentclass[twocolumn]{aastex631}
\usepackage{wrapfig}
\def \grb {\mbox{GRB\, 241107A}}
\def \pgc {\mbox{PGC\, 86046}}

\def \inte {\emph{INTEGRAL}}

\def\ltsima{$\; \buildrel < \over \sim \;$}
\def\lsim{\lower.5ex\hbox{\ltsima}}
\def\gtsima{$\; \buildrel > \over \sim \;$}
\def\gsim{\lower.5ex\hbox{\gtsima}}

\begin{document}

\title{GRB241107A: a Giant Flare from a close-by extragalactic Magnetar?}

\correspondingauthor{James Craig Rodi}
\email{james.rodi@inaf.it}

\author[0000-0003-2126-5908]{James Craig Rodi}
\affiliation{INAF - Istituto di Astrofisica e Planetologia Spaziali di Roma, Via del Fosso del Cavaliere 100, 00133 Roma, Italy}

\author[0009-0001-3911-9266]{Dominik Patryk Pacholski}
\affiliation{INAF - Istituto di Astrofisica Spaziale e Fisica Cosmica di Milano, Via A. Corti 12, 20133 Milano, Italy}
\affiliation{Universit\`a degli Studi di Milano Bicocca, Dipartimento di Fisica G. Occhialini, Piazza della Scienza 3, 20126 Milano, Italy}

\author[0000-0003-3259-7801]{Sandro Mereghetti}
\affiliation{INAF - Istituto di Astrofisica Spaziale e Fisica Cosmica di Milano, Via A. Corti 12, 20133 Milano, Italy}

\author[0009-0001-9705-8019]{Edoardo Arrigoni}
\affiliation{INAF - Istituto di Astrofisica Spaziale e Fisica Cosmica di Milano, Via A. Corti 12, 20133 Milano, Italy}
\affiliation{Universit\`a degli Studi di Milano Bicocca, Dipartimento di Fisica G. Occhialini, Piazza della Scienza 3, 20126 Milano, Italy}

\author[0000-0002-2017-4396]{Angela Bazzano}
\affiliation{INAF - Istituto di Astrofisica e Planetologia Spaziali di Roma, Via del Fosso del Cavaliere 100, 00133 Roma, Italy}

\author[0000-0002-6601-9543]{Lorenzo Natalucci}
\affiliation{INAF - Istituto di Astrofisica e Planetologia Spaziali di Roma, Via del Fosso del Cavaliere 100, 00133 Roma, Italy}

\author[0000-0002-9393-8078]{Ruben Salvaterra}
\affiliation{INAF - Istituto di Astrofisica Spaziale e Fisica Cosmica di Milano, Via A. Corti 12, 20133 Milano, Italy}

\author[0000-0003-0601-0261]{Pietro Ubertini}
\affiliation{INAF - Istituto di Astrofisica e Planetologia Spaziali di Roma, Via del Fosso del Cavaliere 100, 00133 Roma, Italy}




\begin{abstract}
We report the results on the short gamma-ray burst \grb , obtained with the IBIS instrument on board the \inte\ satellite. The burst had a duration of about 0.2 s, a fluence of $8\times10^{-7}$ erg cm$^{-2}$  in the 20 keV-10 MeV range and a hard spectrum, characterized by a peak energy of 680 keV.  The position of \grb\ has been precisely determined because it fell inside the imaging field of view of the IBIS coded mask instrument.  The presence of the nearby galaxy \pgc\ in the 3 arcmin radius error region, suggests that \grb\ might be a giant flare from a magnetar rather than a canonical short GRB. 
For  the 4.1 Mpc distance of \pgc , the  isotropic energy of 1.6$\times10^{45}$ erg is in agreement with this hypothesis, that is also supported by the time resolved spectral properties similar to those of  the few other extragalactic magnetars giant flares detected so far.

\end{abstract}



\section{Introduction}

Since the time of the discovery of the first giant flare from a magnetar on 1979 March 5 \citep{1979Natur.282..587M}, it was realized that these energetic events ($L_{peak}=10^{45-47}$ erg s$^{-1}$) can be detected up to distances of several Mpc by current hard X-ray / $\gamma$-ray instruments \citep{1982Ap&SS..84..173M}. However, the long X-ray tails   periodically modulated by the magnetar rotation that characterize the only three confirmed magnetar giant flares (MGFs), are too faint to be detected from sources located in galaxies farther than the Magellanic Clouds, making such events virtually undistinguishable from short gamma-ray bursts (sGRB).

Despite this difficulty, a few candidate extragalactic MGFs have been identified among short GRBs  positionally consistent with nearby galaxies, often characterized by a high star formation rate \citep{2007AstL...33...19F,2008ApJ...680..545M,2021Natur.589..211S,2021ApJ...907L..28B}.  
The most recent example is 231115A \citep{2024Natur.629...58M}, associated to the starburst galaxy M82 thanks to  a localization with arcmin precision provided in real time by the INTEGRAL Burst Alert System \citep{2003A&A...411L.291M}.

However, the sample of extragalactic MGFs candidates is still rather small, considering that these events are expected to account for a significant fraction of the population of short GRBs, with different estimates   ranging from a few percent  to  a much larger fraction \citep{2005Natur.434.1098H,2006ApJ...640..849N,2007ApJ...659..339O,2015MNRAS.447.1028S,2024arXiv241116846B}. Considering the low rate of giant flares at galactic (Milky Way and Magellanic Clouds) distances (only three observed in 50 years), it is clear that increasing the sample of extragalactic MGFs is the most  promising way to acquire more information on the rate of these events, which, among other things, is relevant to understand the magnetic field evolution of magnetars.

The short \grb\ was first found in data from the $SVOM$ and \inte\ satellites \citep{2024GCN.38125....1S,2024GCN.38164....1R}. An error box with area of about one square degree was obtained by triangulation of data from SVOM, INTEGRAL, Konus-WIND and Swift \citep{2024GCN.38165....1K}.  The only reported optical follow-up observation was performed 4.8 days after the trigger with the GROWTH-India Telescope, which derived upper limits of about 20.8 and 20.6 mag for r\(^\prime\) and i\(^\prime\) filters, respectively \citep{2024GCN.38187....1M}.  Although the burst was inside the fields of view of the INTEGRAL/IBIS and Swift/BAT imaging instruments, it was too faint to trigger the automatic burst searches. The offline analysis of these data provided positions with uncertainties of a few square arcmin \citep{2024GCN.38172....1M,2024GCN.38177....1D} at coordinates consistent with those of the galaxy \pgc . 
%
%
Here we report on the  \inte\ observations of \grb\ and  discuss its  possible  interpretation as a MGF candidate in this nearby galaxy 
($D=4.1^{+1.2}_{-0.9}$ Mpc, 
\citet{2016AJ....152...50T}).

\section{Observations and Data Analysis}

The IBIS imaging telescope \citep{2003A&A...411L.131U}
consists of two detectors operating simultaneously in different energy ranges. The INTEGRAL Soft Gamma-Ray Imager (ISGRI) provides photon-by-photon data in the 15-1000 keV range \citep{2003A&A...411L.141L}, while  the PIxelised CsI Telescope (PICsIT) covers the 175 keV \(-\) 10 MeV range, providing images and light curves integrated in different energy and time bins \citep{2003A&A...411L.149L}.

In the following analysis, we used PICsIT data in spectral-timing mode, which have a time resolution of 3.9 ms with counts integrated over the entire detector in 8 pre-defined energy channels from 212 keV to 2.6 MeV.  Therefore, no imaging information is available for this data type.  The background count rate in each energy channel was taken to be the median count rate during the INTEGRAL pointing (1800 s) including \grb .  To account for the angular distance between \grb\ and the IBIS pointing direction  (11.6 degrees), a correction was applied to the PICsIT effective area.  This procedure is the same that was also used in the analysis of GRB 231115A \citep{2024Natur.629...58M}.

\begin{figure}[htb!]
  \begin{center}
  \includegraphics[width=8.5cm]{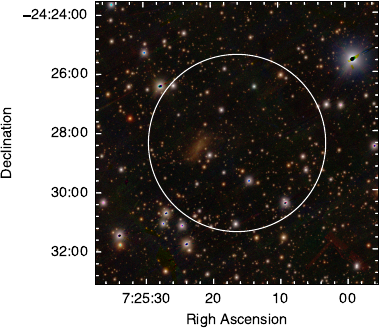}
  \caption{Pan-STARRS1 optical image of the location of \grb\, with the i-band, r-band, and g-band used as RGB colours. The circle with radius 3 arcmin is the IBIS/ISGRI position (90\% c.l.). The galaxy \pgc\ is clearly visible at coordinates R.A.=7:25:22, Dec.=--24:28:23 }  
  \label{fig-ima}
      \end{center}
\end{figure}

\subsection{Position of the burst}

We used version 11.2 of the Off-line Scientific Analysis \citep[OSA,][]{2003A&A...411L.223G} software to extract an image with the ISGRI data (no imaging information with adequate time resolution is provided by PiICsIT).  The burst is detected with the highest significance (6.7$\sigma$) by selecting data in the 30-180 keV energy range and in the time interval from T$_0+$0.17 s to T$_0$+0.40 s  (with T$_0$=2024-11-07 23:30:00 UTC).  The derived coordinates are R.A.= 111.3360 deg,  Dec.= $-$24.4439 deg (J2000) with an uncertainty of 3 arcmin (90\% c.l. radius). This position is consistent with, and supersedes, the one derived using the preliminary satellite attitude information \citep{2024GCN.38172....1M}.
Fig.~\ref{fig-ima}  shows the error region of \grb\ superimposed on an optical image from  Pan-STARRS1 \citep{2016arXiv161205560C}, where the \pgc\ galaxy is clearly visible.

 \begin{figure}[hb]
  \begin{center}
  \includegraphics[]{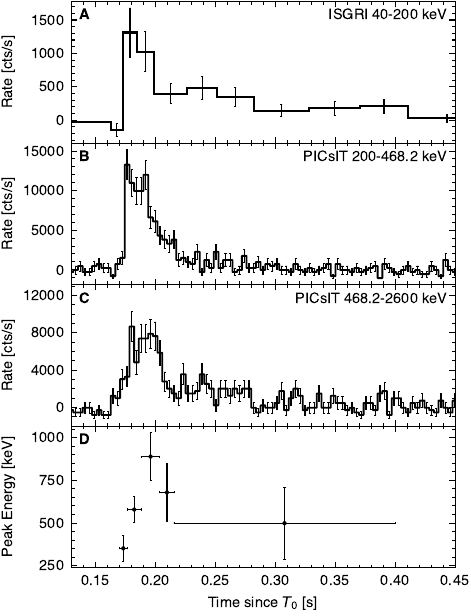}
  \caption{Background subtracted light curves of \grb\ in three   energy bands, with $T_0$=2024-11-07 23:30:00  UT. The ISGRI (40-200 keV) light curve  has been binned to have at least 18 counts in each bin (A panel). The  PICsIT light curves (200-468.2 keV, B panel and 468.2-2600 keV, C panel) have the original bin size of 3.9 ms. Panel D shows the evolution of $E_p$ during the burst.}  
  \label{fig-lc}
  \label{fig:lc_spectrum}
      \end{center}
\end{figure}

\begin{table*}
\begin{center}
\caption{Results of Spectral Fits}
 \hspace{-30mm}
\hspace{-6mm}
\begin{tabular}{cccccccc}
\tableline
Interval & Start-Stop$^a$ & Model$^b$ &  $\alpha$ & $E_p$ & $T_{BB}$ & Flux$^c$ &  \(\chi^2\)(dof)  \\
         & [s]       &       &           &  [keV] & [keV] & $[10^{-6}$ erg cm$^{-2}$ s$^{-1}$] &    \\
\tableline
Total & 0.17--0.40 & CPL & $0.07^{+0.27}_{-0.24}$ & $678^{+125}_{-90}$ & -- & $3.4\pm0.5$ & 8.8(9) \\
      &            & BB &  -- & --  & $136^{+15}_{-11}$ & $2.7^{+0.4}_{-0.2}$ & 21.8(10) \\
\hline
Peak  & 0.17--0.20 & CPL & $0.71^{+0.42}_{-0.31}$ & $674^{+85}_{-71}$ & -- & $16.1^{+1.9}_{-1.8}$ & 14.3(9) \\
      &            & BB  & --  & -- &  $152^{+18}_{-9}$ & $14.7^{+1.4}_{-1.5}$ & 17.0(10) \\
Tail & 0.20--0.40 & CPL & $-0.02^{+0.48}_{-0.39}$ & $460^{+126}_{-88}$ & -- & $1.1\pm0.2$   & 13.9(8)   \\
      &            & BB  & -- & -- & $99^{+22}_{-20}$ &  $1.0\pm0.2$ & 19.3(9) \\
\tableline
1 &  0.1700\(-\)0.1765 & CPL & 0.07$^d$ & \(352\pm76\)   & &  \(8^{+2}_{-2} \)       & 4.75(5) \\
2 &  0.1765\(-\)0.1882 & CPL & 0.07$^d$ & \(579\pm76\)   & &  \(24^{+5}_{-4}\) & 9.80(5) \\ 
3 &  0.1882\(-\)0.2038 & CPL & 0.07$^d$ & \(889\pm140\) & &  \(25^{+8}_{-6} \) & 5.00(5) \\ 
4 &  0.2038\(-\)0.2155 & CPL & 0.07$^d$ & \(679\pm172\) & &  \(9^{+3}_{-2}\)       & 3.10(5) \\
5 &  0.2155\(-\)0.4000 & CPL & 0.07$^d$ & \(499\pm211\) & &  \(0.6^{+0.4}_{-0.3}\)       & 4.36(4) \\
\tableline
\end{tabular}
\label{table:pspectra_fits}
\end{center}
\tablecomments{
\vspace{-0.25cm}
\tablenotetext{a}{Times referred to T$_0$}\vspace{-0.25cm}
\tablenotetext{b}{CPL = cutoff power law, BB = blackbody}\vspace{-0.25cm}
\tablenotetext{c}{in the 20 keV -- 10 MeV  energy range}\vspace{-0.25cm}
\tablenotetext{d}{fixed}
}
\end{table*}

\subsection{Timing and spectral properties}
\label{sec:tim_sp}
 
The light curves of \grb\ as measured by IBIS in different energy ranges are plotted in Fig.~\ref{fig-lc}. The burst  had a total duration of $\sim$250 ms and consisted of an initial  pulse lasting $\sim$50 ms followed by a fainter tail. An image extracted from the ISGRI data in the time interval from T$_0+$0.25 s to T$_0$+0.40 s confirms that the tail is indeed due to the burst.
Based on the PICsIT light curve, the GRB rise occurred within a single time bin and thus it lasted less than 3.9 ms.  

\begin{figure}[ht]
  \begin{center}
  \includegraphics[scale=0.5, angle=0,trim = 20mm 5mm 10mm 100mm, clip]{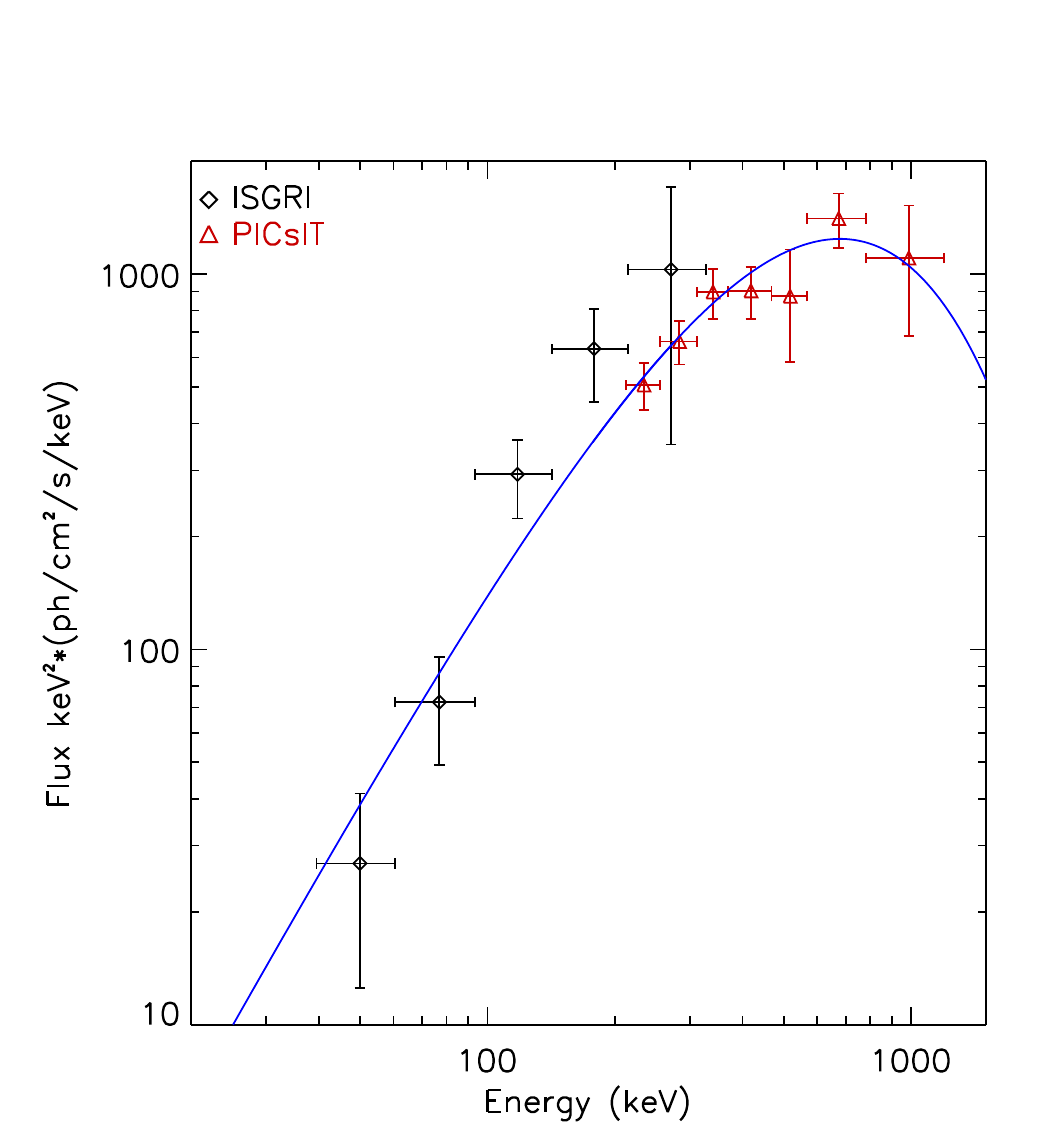}
  \caption{Time averaged IBIS spectrum of \grb\ (from T$_0$+0.17 s to T$_0$+0.40 s).
  ISGRI data are plotted as black diamonds and PICsIT data as red triangles.  The best fit cutoff power-law model is overplotted as a solid blue line.} 
  \label{fig:avg_spectrum}
      \end{center}
\end{figure}

The time averaged spectrum of \grb , obtained from the ISGRI and PICsIT  data of the whole duration of the burst, is shown in Fig.~\ref{fig:avg_spectrum}. 
A good fit is found  with an exponentially cut off power law  model 
with photon index \(\alpha = 0.07^{+0.27}_{-0.24}\),  peak energy \(E_p = 678^{+125}_{+90}\) keV
and  20 keV \(-\) 10 MeV flux of (3.4$\pm$0.5)$\times10^{-6}$ erg cm$^{-2}$ s$^{-1}$. 
This corresponds to a fluence of  $\sim8\times10^{-7}$ erg cm$^{-2}$. 
A   single blackbody model gives   a temperature of \(kT = 136^{+15}_{-11}\) keV, but the fit is worse. 
All the fit parameters are given in Table~\ref{table:pspectra_fits}.   \\



\begin{figure}[hb]
  \begin{center}
  \includegraphics[]{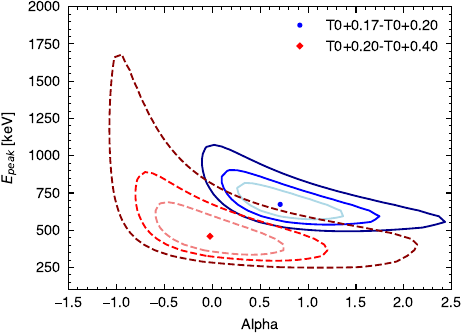}
  \caption{Confidence contours of the joint ISGRI+PICsIT spectra for the peak (T$_0$+0.17-T$_0$+0.20, solid blue line) and tail (T$_0$+0.20-T$_0$+0.40, dashed red line) of the burst. The contours represent confidence levels of 68\%, 90\%, and 99\%.} 
  \label{fig:contour}
      \end{center}
\end{figure}

We also extracted ISGRI and PICsIT spectra for two time intervals corresponding to the main pulse and the tail. Although the best fit parameters have relatively large uncertainties, they seem to indicate that the spectrum of the tail is slightly softer than that of the peak (see Table~\ref{table:pspectra_fits}). As it is shown by the contour plots of the errors on $\alpha$ and $E_p$  shown in Fig.~\ref{fig:contour}, the spectral variation can be described by a reduction in the peak energy, rather than by a change in $\alpha$.

To further investigate the burst spectral evolution we split the PICsIT data into the five time intervals indicated in Table~\ref{table:pspectra_fits}
and performed a joint fit to an exponentially cutoff power law with \(\alpha\) fixed to the time-averaged value (0.07). 
We found that at the start of the burst \(E_p\) increases from \(\sim 330\) keV to a maximum value of \(\sim 890\) keV at the peak before decreasing to roughly 500 keV as the flux decays though the errors on the \(E_p\) values are large (see bottom panel of  Figure~\ref{fig:lc_spectrum} and Table~\ref{table:pspectra_fits}). 

The sky position of \grb\ was repeatedly observed by \inte\ starting from February 2003 to $T_0+5.7$ days. We extracted from the public data archive all the relevant IBIS pointings, totalling 9 Ms of exposure. Using the method described in \citet{2021ApJ...921L...3M} and \citet{2024MNRAS.tmp.2449P},  we searched for other possibile bursts from this position,  but none was found.

\section{Discussion and conclusions}

The short duration and hard spectrum of \grb\ are consistent with the properties of short GRBs, but its possible association with a nearby galaxy leads us to also consider the alternative interpretation in terms of a MGF.

The chance coincidence of finding a galaxy brighter than magnitude $m$ in an error circle of radius $R$ is given by
\begin{equation}
P(<m)= 1-\mathrm e^{-\pi R^2 \sigma(<m)}
\end{equation}
where, $\sigma(<m)$ is the number density of galaxies brighter than $m$. We derived $\sigma(<m)$ from the galaxy number counts  plotted in Fig.~4 of \cite{2000ARA&A..38..667F} and used the  \pgc\ magnitudes in the I and B bands, $m_B = 16.336,\ m_I = 14.634$, reported in the HIPASS catalogue \citep{2005MNRAS.361...34D}. For $R$=3 arcmin, we obtain  chance coincidence probabilities of $P(<m_B)$=2.9\% and $P(<m_I)$=5.5\%.

\begin{figure}[ht]
  \begin{center}
  \includegraphics[width=8cm]{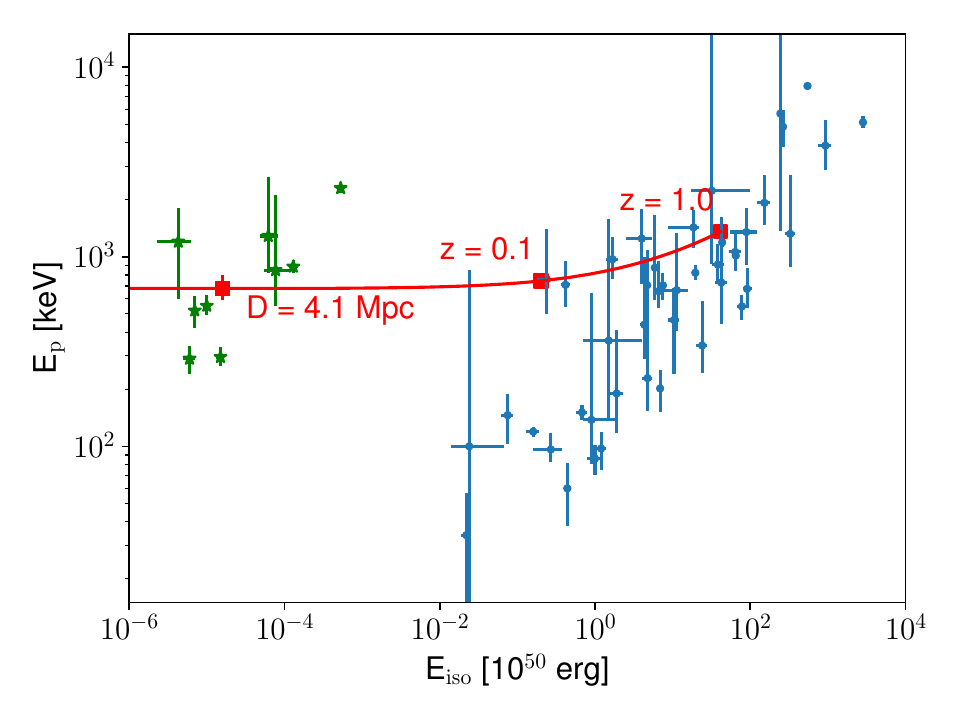}
  \caption{Position of \grb\ (red square) in the $E_p$ versus $E_{iso}$ plane. The sample of short GRBs  (blue) is taken from \citet{2020MNRAS.492.1919M}. The three confirmed magnetar giant flares and the extragalactic MGF candidates are indicated by the green stars   (data from \cite{2024MNRAS.tmp.2449P} and references therein and \cite{2016ApJS..224...10S}).}
  \label{fig:EpEiso}
      \end{center}
\end{figure}

At the \pgc\ distance of 4.1 Mpc \citep{2016AJ....152...50T}, the fluence derived in Sec.~\ref{sec:tim_sp} corresponds to an emitted isotropic energy $E_{iso}=1.6\times10^{45}$ erg, that fits perfectly with the typical values of MGFs. 
This is illustrated in the $E_p-E_{iso}$ plot of Fig.~\ref{fig:EpEiso}, where  the values of \grb\ for different assumed distances are  compared with those of the other MGFs and of the short GRBs.  

It is also interesting to compare \grb\ with the MGF candidate in the Sculptor Galaxy,  GRB 200415A \citep{2021Natur.589..211S,2021Natur.589..207R}, which, in the small sample of extragalactic MGFs, is probably the one with the best spectral information. During  the initial $\sim$7 ms, its spectrum hardened, with $E_p$ evolving from 430 keV to 1.8 MeV, and then it gradually softened with $E_p$ tracing the downward flux evolution.  A similar behavior is possibly seen also in GRB 231115A \citep{2024arXiv240906056T}, which is currently the most convincing extragalactic MGF and is associated to the M82 starburst galaxy \citep{2024Natur.629...58M}. The spectral evolution of \grb\ shown in Fig.~\ref{fig:lc_spectrum} is consistent with a similar behavior. These three bursts are also similar for their rise times $\lesssim$4 ms and some evidence for a double-peaked light curve. 

For GRB 200415A time-resolved analysis of the relationship between \(E_p\) and the isotropic luminosity (\(L_{iso}\)) found \(E_p \propto L_{iso}^{0.23 \pm 0.10}\) \citep{2021RAA....21..236C}.  The authors report a stronger correlation with an exponent of \(0.31 \pm 0.04\) when excluding the three spectra before \(T_0 - 0.001\)s.  In the case of GRB 231115A, the exponent is \(0.35^{+0.11}_{-0.07}\) or \(0.41^{+0.21}_{-0.08}\) depending on the time binning \citep{2024arXiv240906056T}.  A fit to our values for GRB 241107A has an exponent of \(0.32 \pm 0.15\), similar to that of the other two MGF candidates.

In conclusion, although we cannot exclude that \grb\ is an ordinary short GRB at redshift $\gtrsim$0.1, its properties are consistent with a MGF origin, as it is suggested by the presence of a nearby galaxy with a chance probability of only a few percent of being in the burst error region. Unfortunately, \grb\ could not be localized precisely in near real time. Rapid follow-up observations at other wavelengths could have provided compelling evidence in favor of one of the two possibilities for the nature of \grb .

\begin{acknowledgments}

The authors thank the referee for the helpful comments.  The  results reported in this article are based on data obtained
with INTEGRAL, an ESA mission with instruments and science data centres funded by ESA member states, and with the participation of the Russian Federation and the USA.
This work received financial support from INAF through the Magnetars Large Program Grant (PI S.Mereghetti). The authors thank the Italian Space Agency for the financial support under the “INTEGRAL ASI-INAF” agreement n◦ 2019-35-HH.0. 
The research leading to these results has received funding from the European Union’s Horizon 2020 Programme under the AHEAD2020 project (grant agreement n. 871158) (J. Rodi).

\end{acknowledgments}

%
\vspace{5mm}






\bibliography{MGF}
\bibliographystyle{aasjournal}



\end{document}